\documentstyle[aps,preprint,epsf]{revtex}

\begin{document}
\draft
\tightenlines

\title{\bf 
Superconductivity in an Exactly Solvable Model of the Pseudogap State:
Absence of Self-Averaging.}
\author{E.Z.Kuchinskii,\ M.V.Sadovskii}
\address
{Institute for Electrophysics, \\
Russian Academy of Sciences,\ Ural Branch, \\ 
Ekaterinburg,\ 620016, Russia\\
E-mail:\ kuchinsk@iep.uran.ru,\ sadovski@iep.uran.ru} 
\maketitle


\begin{abstract}
We analyze the anomalies of superconducting state within a simple exactly
solvable model of the pseudogap state, induced by fluctuations of ``dielectric''
short range order, for the model of the Fermi surface with ``hot'' patches.
The analysis is performed for the arbitrary values of the correlation length
$\xi_{corr}$ of this short range order. It is shown that superconducting
energy gap averaged over these fluctuations is non zero in a wide temperature
range above  $T_c$ --- the temperature of homogeneous superconducting
transition. This follows from the absence of self averaging of
the gap over the random field of fluctuations. For temperatures $T>T_c$ 
superconductivity apparently appears in separate regions of space (``drops'').
These effects become weaker for shorter correlation lengths $\xi_{corr}$
and the region of ``drops'' on the phase diagram becomes narrower and
disappears for $\xi_{corr}\to 0$, however, for the finite values of
$\xi_{corr}$ the complete self averaging is absent. 
\end{abstract} 

\newpage
\narrowtext

\section{Introduction}

Among the anomalies of electronic properties of high -- temperature
superconducting cuprates (HTSC) especially interesting are the properties of
the pseudogap state, observed in a wide region of the phase diagram 
\cite{Tim,MS}. From our point of view, the preferential scenario for the
formation of the pseudogap state is based on the existence (mainly in the
underdoped region of the phase diagram) of strong scattering of current
carriers on well developed fluctuations of ``dielectric'' short range
order (of antiferromagnetic (AFM) or charge density wave (CDW) type) \cite{MS}.
This scattering leads to essentially non -- Fermi liquid like renormalization
of electronic spectrum in certain parts of momentum space close to the Fermi
surface and near the so called ``hot'' spots or ``hot'' (flat) patches 
\cite{MS}. The preferential nature of ``dielectric'' and not of
``superconducting'' pseudogap formation \cite{Lok} follows from the number
experiments, the appropriate discussion can be found in Ref. \cite{MS}.

The major part of existing theoretical papers is dealing with pseudogap
formation and its influence on the properties of the system in normal
(non superconducting) state, while only few are consider the anomalies of
superconductivity in the pseudogap state \cite{PS,KS,KS1}. In particular,
in Ref. \cite{KS} we have analyzed superconductivity in a simple exactly
solvable model of the pseudogap state, based on the model Fermi surface (in
two dimensions) with ``hot'' patches \cite{PS}. For the description of the
pseudogap state we have used an exactly solvable model first developed for the
one dimensional case in Ref. \cite{S74} and for the limit of very large
correlation lengths of ``dielectric'' short range order. It was shown that
the superconducting energy gap averaged over the fluctuations of short range
order is, in general, non zero also in the temperature region above
the ``mean-field'' temperature of superconducting transition $T_c$, 
corresponding (according to Ref. \cite{KS}) to the appearance of homogeneous
superconducting state in a sample as a whole. This lead us to the conclusion
\cite{KS}, that for temperatures $T>T_c$ there appear superconducting
``drops'', which exist up to the temperature $T_{c0}$ of superconducting
transition in the absence of the pseudogap of ``dielectric'' nature.
This effect was attributed in Ref. \cite{KS} to the absence of self averaging
of superconducting order parameter (gap) in situation when the correlation
length of short range order fluctuations is larger than the coherence length
of superconductivity (the size of Cooper pairs).

Under the assumption of self averaging superconducting energy gap the effects
of finiteness of the correlation length of short range order fluctuations
were analyzed in Ref. \cite{KS1}, where we have considered the pseudogap
influence on $T_c$, calculated the temperature dependence of the gap for
$T<T_c$, and derived the Ginzburg -- Landau expansion for $T\sim T_c$. 
In this case we have used our ``nearly exact'' solution for the pseudogap
state induced by Gaussian fluctuations of short range order, proposed first
in Refs. \cite{S79,S91} for one dimensional case and generalized for two
dimensions in Refs. \cite{Sch,KS2}. Within this approach it seems rather 
difficult to find the solution without the assumption of the self averaging
superconducting energy gap. Note that the problem of the presence or absence of
self averaging of superconducting gap was not studied well enough in most of the
previous work on disordered superconductors. In most studies the self
averaging property was just assumed on ``physical grounds'', with the 
reference on significantly different length scales on which the superconducting
gap changes (coherence length $\xi_0$) and characteristic scales for the
electronic subsystem (interatomic distance or the inverse Fermi momentum), e.g.
in the impurity problem  \cite{Gor,Genn,SL}, or the correlation length  
$\xi_{corr}$ of short range order in our problem \cite{MS,KS,KS1}), where it
may be expected that the complete self averaging of the superconducting
energy gap appears for  $\xi_{corr}\ll\xi_0$ \cite{MS,KS1}. We are unaware of
any work, where the problem of self averageness was analyzed within some
exactly solvable model of disorder.

One of the main aims of the present work is to perform precisely this type
of analysis within very simple (though probably not realistic enough) 
one dimensional model of the pseudogap state, induced by fluctuations of 
``dielectric'' short range order with finite correlation length, proposed in
a recent paper of Bartosch and Kopietz \cite{BK}. An exact solution found in 
this work is very similar to that of our previous studies \cite{S74,S79,S91}
and allows to perform a complete analysis of the self averaging properties of
superconducting energy gap for two dimensional model of ``hot'' patches,
proposed in Refs. \cite{PS,KS1,KS2}. Besides we shall study the full temperature
dependences of superconducting energy gap for a superconductor with the
pseudogap of ``dielectric'' nature.

\section{Simplified Model of the Pseudogap State.}

Let us consider an exactly solvable model of the pseudogap state proposed in
Ref. \cite{BK} using slightly different approach. Consider an electron
in one dimension in a periodic potential of the following form:
\begin{equation}
V(x)=2D\cos(Qx+\phi)
\label{Vx}
\end{equation}
with $Q=2p_F-k$, where $p_F$ -- is Fermi momentum, and $k\ll p_F$ -- is some
small deviation from the scattering vector $2p_F$ \footnote{This choice of
the vector of AFM or CDW superstructure corresponds, in general, to the case
of incommensurate ordering and fluctuations.}. Electronic spectrum close to
the Fermi level is taken in the usual linearized form:
\begin{eqnarray}
\xi_1\equiv\xi_p=v_F(|p|-p_F)\qquad \xi_{p-2p_F}=-\xi_p\quad
\mbox{(``нестинг'')}\nonumber\\
\xi_2\equiv\xi_{p-Q}=-\xi_p-v_Fk\equiv -\xi_p-\eta
\label{el-sp}
\end{eqnarray}
where we introduced the notation $\eta=v_Fk$ ($v_F$ -- Fermi velocity), which
will be widely used in the following. The potential (\ref{Vx}) can be rewritten
as:
\begin{equation}
V(x)=De^{i2p_Fx-ikx}+D^*e^{-i2p_Fx+ikx}
\label{Vxc}
\end{equation}
where we have introduced the complex amplitude $D\to De^{i\phi}$.

The solution of this problem is elementary.
 In the ``two-wave'' approximation
of the usual band theory the one-electron (normal) Green's function, 
corresponding to the diagonal transition $p\to p$, takes the following
form (in Matsubara's representation):
\begin{eqnarray}
g_{11}(i\varepsilon_n pp)=\frac{1}{i\varepsilon_n-\xi_1}+
\frac{1}{i\varepsilon_n-\xi_1}D^*\frac{1}{i\varepsilon_n-\xi_2}D
\frac{1}{i\varepsilon_n-\xi_1}+...=\nonumber\\
=\frac{i\varepsilon_n-\xi_2}{(i\varepsilon_n-\xi_1)(i\varepsilon_n-\xi_2)-
|D|^2}=\frac{i\varepsilon+\xi+\eta}{(i\varepsilon-\xi)(i\varepsilon+\xi+\eta)-
|D|^2}
\label{g11}
\end{eqnarray}
where in the last equality we have introduced the following short notations:
$\xi_p=\xi$, $\varepsilon_n=\varepsilon$. We can also introduce the non
diagonal (anomalous) Green's function, corresponding to the Umklapp process
$p\to p-Q$:
\begin{eqnarray}
g_{12}(i\varepsilon_n pp-Q)=
\frac{1}{i\varepsilon_n-\xi_1}D^*\frac{1}{i\varepsilon_n-\xi_2}+...=\nonumber\\
=\frac{D^*}{(i\varepsilon_n-\xi_1)(i\varepsilon_n-\xi_2)-
|D|^2}=\frac{D^*}{(i\varepsilon-\xi)(i\varepsilon+\xi+\eta)-
|D|^2}
\label{g12}
\end{eqnarray}

Let now (\ref{Vx}) be random field. Following  Ref. \cite{BK} we shall
consider the very special model of disorder, when we shall assume the
randomness of the scattering vector deviation $k$, with Lorentzian distribution
\footnote{In fact this corresponds to a specific model of phase fluctuations
of the potential (\ref{Vx}).}:
\begin{equation}
{\cal P}_{k}(k)=\frac{1}{\pi}\frac{\kappa}{k^2+\kappa^2}
\label{Lork}
\end{equation}
where $\kappa\equiv\xi_{corr}^{-1}$ and $\xi_{corr}$ -- is the correlation 
length of short range order.
Phase $\phi$ in (\ref{Vx}) is also considered as random with uniform
distribution between $0$ and $2\pi$:
\begin{equation}
{\cal P}_{\phi}(\phi)=\left\{
\begin{array}{l}
\frac{1}{2\pi}\qquad\mbox{for}\quad 0\leq\phi\leq 2\pi\\
0\qquad\mbox{for other values}
\end{array}
\right.
\label{Pfi}
\end{equation}
Correlation function of random fields $V(x)$ at different points can be
calculated directly and is given by:
\begin{equation}
<V(x)V(x')>=2D^2\cos[2p_F(x-x')]\exp[-\kappa|x-x'|]
\label{VxVx}
\end{equation}
where the angular brackets denote averaging over (\ref{Lork}) and (\ref{Pfi}).
The random field with precisely this form of pair correlator was treated first
in Ref. \cite{LRA}, as well as in Refs. \cite{S74,S79,S91}, though in these
works the Gaussian statistics of the random field was also assumed
\footnote{For the Gaussian random field all higher correlators of the random
field $V(x)$ are factorized ``a'la Wick'' into products of pair correlators 
(\ref{VxVx}).}. In the model under consideration the random field
$V(x)$ is, in general, non Gaussian \cite{BK}.
The Fourier transformation of (\ref{VxVx}) takes the form of characteristic 
Lorentzian, which defines the effective interaction of an electron with
fluctuations of short range order \cite{MS}:
\begin{equation}
V_{eff}(q)=2D^2\left\{\frac{\kappa}{(q-2p_F)^2+\kappa^2}+
\frac{\kappa}{(q+2p_F)^2+\kappa^2}\right\}
\label{Veff}
\end{equation}
This type of effective interaction was introduced in all papers on the
pseudogap of ``dielectric'' nature cited above.

Green's functions averaged over the ensemble of random fields (\ref{Vx}) with
distributions (\ref{Lork}) and (\ref{Pfi}) are calculated by elementary
integration. The average value of the anomalous Green's function (\ref{g12})
is simply zero (after the averaging over (\ref{Pfi})), which corresponds to
the absence of ``dielectric'' long range order. The average of the Green's
function (\ref{g11}) is obtained by term by term integration of the
perturbation series (\ref{g11}) with (\ref{Lork}), so that:
\begin{eqnarray}
G(i\varepsilon_n p)=\frac{1}{i\varepsilon_n-\xi_p}+
\frac{1}{i\varepsilon_n-\xi_p}D^*\frac{1}{i\varepsilon_n+\xi_p+iv_F\kappa}D
\frac{1}{i\varepsilon_n-\xi_p}+\nonumber\\
+\frac{1}{i\varepsilon_n-\xi_p}D^*\frac{1}{i\varepsilon_n+\xi_p+iv_F\kappa}D
\frac{1}{i\varepsilon_n-\xi_p}D^*\frac{1}{i\varepsilon_n+\xi_p+iv_F\kappa}D
\frac{1}{i\varepsilon_n-\xi_p}+...=\nonumber\\
=\frac{i\varepsilon_n+\xi_p+iv_F\kappa}
{(i\varepsilon_n-\xi_p)(i\varepsilon_n+\xi_p+iv_F\kappa)-|D|^2}
\label{Gp}
\end{eqnarray}
This is precisely the exact solution for the Green's function given in
Ref. \cite{BK}.

In the following we shall also consider the model with fluctuating amplitude
$D$ of the field (\ref{Vx}), so that the appropriate Green's function can be
obtained by the averaging of (\ref{Gp}) with some amplitude distribution
${\cal P}_{D}(D)$.
 In particular, we can take the amplitude distribution in
the Rayleigh form \cite{S74,S79,BK}:  
\begin{equation} 
{\cal P}_{D}(D)=\frac{2D}{W^2}\exp\left(-\frac{D^2}{W^2}\right) 
\label{Rayl} 
\end{equation}
In this case the additional averaging of correlators (\ref{VxVx}) and 
(\ref{Veff}) leads to the simple replacement $D\to W$. The average Green's
function of an electron in this case becomes:
\begin{eqnarray}
G(i\varepsilon_n p)=\int_{0}^{\infty}dD{\cal P}_D(D)
\frac{i\varepsilon_n+\xi_p+iv_F\kappa}
{(i\varepsilon_n-\xi_p)(i\varepsilon_n+\xi_p+iv_F\kappa)-|D|^2}=\nonumber\\
=\int_{0}^{\infty}d\zeta e^{-\zeta}
\frac{i\varepsilon_n+\xi_p+iv_F\kappa}
{(i\varepsilon_n-\xi_p)(i\varepsilon_n+\xi_p+iv_F\kappa)-\zeta W^2}
\label{GpR}
\end{eqnarray}
where $W$ determines now the effective width of the pseudogap. In the limit of
the large correlation length of fluctuations of (\ref{Vx}), i.e. for
$\xi_{corr}\to\infty$
 ($\kappa\to 0$), the solution (\ref{GpR}) coincides with
that found earlier in Refs. \cite{S74} for the case of the Gaussian random
field. For finite $\kappa$ this solution coincides with that proposed in
Ref. \cite{KS2} during the formal analysis of approximations, used in
Refs. \cite{S79,S91} in the analysis of the general problem of an
electron in the Gaussian random field with pair correlator given by
(\ref{VxVx}). In Refs. \cite{KS2,BK} it was shown that the density of states,
corresponding to the Green's function (\ref{GpR}), possesses a characteristic
smooth pseudogap in the vicinity of the Fermi level, and the values of the
density of states are quantitatively very close \cite{KS2,BK,Sad}
(practically for all energies in the incommensurate case) to the values
obtained in Ref. \cite{S79}, as well as to the results of exact numerical
calculation for the Gaussian random field, performed in Refs. \cite{B1,B2,MM}
\footnote{Using the approach of Ref. \cite{S74} in the present model it is easy
to find also the two particle Green's function and accordingly the frequency
dependence of conductivity \cite{BK}. Unfortunately, the specific form of
``disorder'' (random field) leads to unphysical behavior at zero frequency,
corresponding to an ``ideal'' conductor.}.

If the random field (\ref{Vx}) is created by fluctuations of some kind of
``dielectric'' order parameter (e.g. CDW or AFM), distribution (\ref{Rayl})
may describe Gaussian fluctuations, existing at high enough temperatures
\cite{Sch,KS2}. For lower temperatures, even before the appropriate long range 
order appears, the amplitude fluctuations of the order parameter are ``frozen 
out'' (cf. \cite{Lok,BD}) and we may assume the amplitude to be non random and
put $D=W$, while phase fluctuations remain important up to much lower
temperatures. Accordingly, the solution of the form of (\ref{Gp}), leading to
sharp enough pseudogap for large correlation lengths $\xi_{corr}$ \cite{LRA}, 
can be used to describe the low temperature region of fluctuations of short
range order. As we do not analyze the microscopics of ``dielectric''
fluctuations, all the parameters, characterizing these fluctuations, such as 
correlation length $\xi_{corr}=\kappa^{-1}$ and amplitudes $D$ and $W$ 
(the energy width of the pseudogap) are considered as phenomenological.
The ``low temperature'' and ``high temperature'' regime of fluctuations of
short range order can, in principle, be realized at different temperatures in
comparison to the temperature of superconducting transition.

Generalization to the case of two dimensional system of electrons, typical for
HTSC -- cuprates, can be done in the spirit of ``hot patches'' model for the
Fermi surface, proposed in Refs. \cite{PS,KS,KS1}. In this case we shall assume
the existence of two independent types of fluctuations of the type of
(\ref{Vx})
\footnote{Note the crude analogy of this picture to the concept of
phase separation in HTSC -- cuprates (stripes) \cite{Str}, if we associate
the correlation length $\xi_{corr}$ with characteristic size (period) of
stripes \cite{MS}.}, oriented along the orthogonal axes  $x$ and $y$, 
strongly
interacting only with electrons from flat parts of the two dimensional Fermi
surface, orthogonal to these axes. Accordingly, we assume the factorized
form of two dimensional (random) potential for electrons:
$V(x,y)=V(x)V(y)$ \cite{PS,KS,KS1}. The size of these flat (``hot'') parts of
the Fermi surface is determined by an additional parameter 
$\alpha$, 
so that $2\alpha$ 
gives the angular size of the flat part, as seen from
the center of the Brillouin zone  \cite{MS,PS,KS,KS1}. In particular, the
value of $\alpha=\pi/4$ corresponds to the square like Fermi surface (complete
nesting), when all Fermi surface becomes ``hot''. For $\alpha<\pi/4$ the
``cold'' parts of the Fermi surface appear, where we neglect the scattering on
fluctuations of the ``dielectric'' order parameter and electrons are considered
as free. In this model all the physical characteristics, determined by the
integrals over the Fermi surface, consist of additive contributions from
``hot'' and ``cold'' parts. Pseudogap renormalization of electronic spectrum
takes place only on ``hot'' parts (and close to these parts), while on
``cold'' part the usual Fermi liquid (gas) behavior remains \cite{MS}. 

This picture is in qualitative agreement with a numerous ARPES experiments
on underdoped cuprates \cite{Tim,MS}, which show that pseudogap anomalies
appear in the vicinity of 
$(0,\pi)$ point of the Brillouin zone, and vanish
as we move to the diagonal direction. The presence of flat parts on the Fermi
surface of HTSC -- cuprates was also reliably observed in the ARPES experiments
by several independent groups \cite{MS}.

\section{Gor'kov's equations in the Pseudogap State.}

To study superconductivity for the system with pseudogap due to 
fluctuations of ``dielectric'' short range order we shall assume the simplest
BCS form of pairing interaction, characterized by the attraction constant  $V$,
which is non zero in some energy interval $2\omega_c$ in the vicinity of the
Fermi level ($\omega_c$ - is the characteristic frequency of the quanta,
responsible for electron attraction). We have already used the same approach
in Refs. \cite{PS,KS,KS1}.
 In the present work we shall limit ourselves only to
the analysis of $s$-wave pairing. There are no serious obstacles within our approach
for the analysis of $d$-wave pairing, typical for HTSC -- cuprates, though in
this case the presence of the angular dependence (anisotropy) of superconducting
gap leads \cite{PS,KS} to the additional integration over the angle and to
a significant growth of time necessary for numerical computations. At the same
time, it was shown in Refs. \cite{PS,KS,KS1}
, that the pseudogap influence on
superconductivity is essentially similar both for $s$ and $d$-wave cases,
differing only by the scale of parameters leading to the same changes of basic
characteristics of superconducting state ($d$-wave pairing is less stable to the
dielectrization of electronic spectrum in comparison to $s$-wave pairing).

\begin{figure}
\epsfxsize=16cm
\epsfysize=12cm
\epsfbox{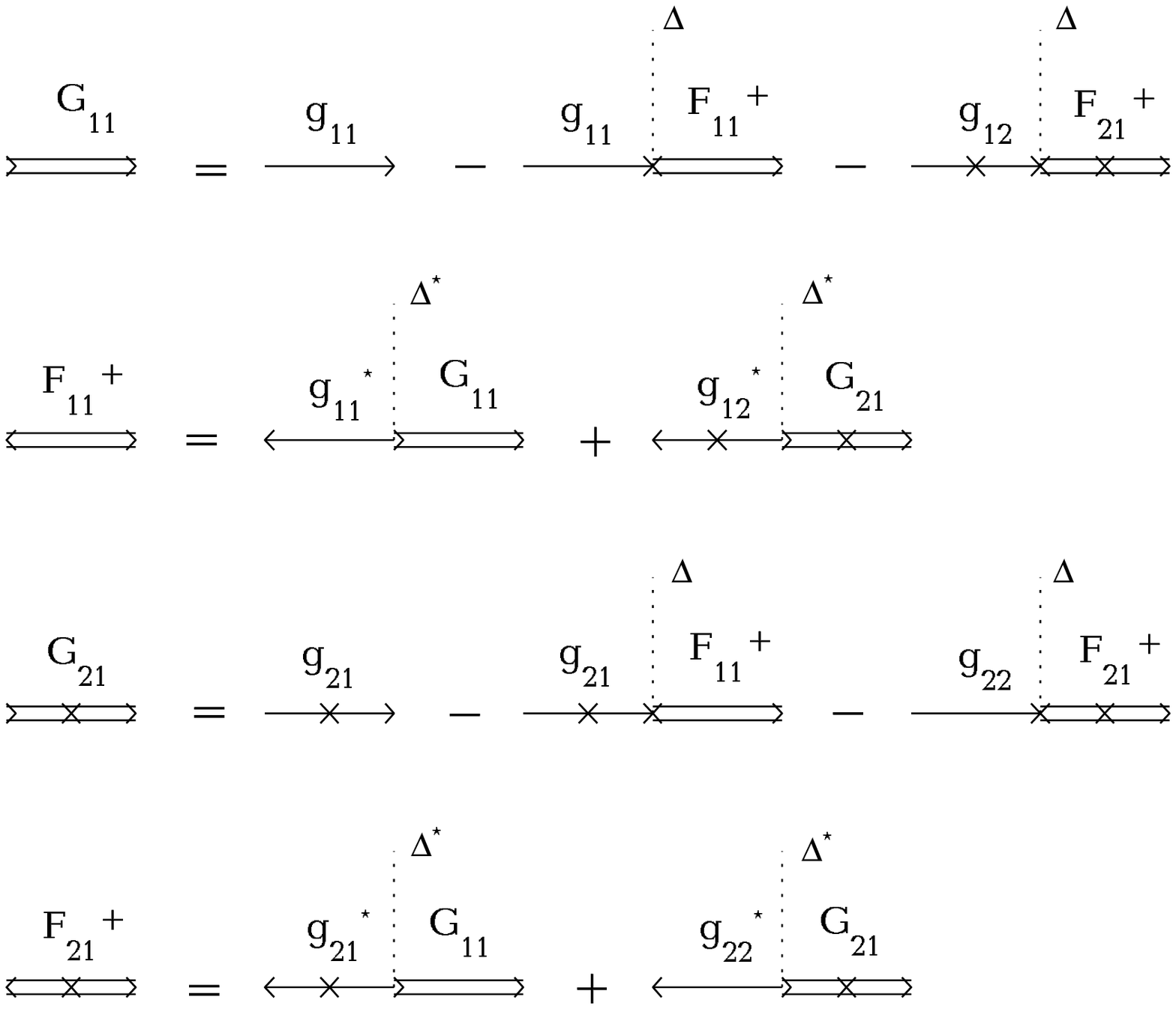}
\caption{Gor'kov's equations in one dimensional periodic field.}
\label{figdiag}
\end{figure}

On ``cold'' parts of the Fermi surface superconductivity is described by
standard BCS equations. In the following we shall concentrate on the derivation
of Gor'kov's equations for the one dimensional model, which is equivalent
to the analysis of the situation at ``hot'' parts of the Fermi surface for 
two dimensional case \cite{KS,KS1}. Green's functions (\ref{g11}), (\ref{g12}) 
for one dimensional system in periodic field (\ref{Vx}),
 can be written as
matrix elements:
\begin{eqnarray}
g_{11}=\frac{i\varepsilon_n-\xi_2}{(i\varepsilon_n-\xi_1)(i\varepsilon_n-\xi_2)-
|D|^2}\qquad
g_{12}=\frac{D^*}{(i\varepsilon_n-\xi_1)(i\varepsilon_n-\xi_2)-|D|^2}
\nonumber \\
g_{21}=\frac{D}{(i\varepsilon_n-\xi_1)(i\varepsilon_n-\xi_1)-|D|^2}\qquad
g_{22}=\frac{i\varepsilon_n-\xi_1}{(i\varepsilon_n-\xi_1)(i\varepsilon_n-\xi_2)-
|D|^2}
\label{GVx}
\end{eqnarray}
In the presence of Cooper pairing the Gor'kov's equations, constructed on
``free'' Green's functions of the type of (\ref{GVx}), are shown in
diagrammatic form in Fig. \ref{figdiag}. In analytic form we have:
\begin{eqnarray}
G_{11}=g_{11}-g_{11}\Delta F^+_{11}-g_{12}\Delta F^+_{21}\nonumber\\
F^+_{11}=g_{11}^*\Delta^*G_{11}+g^*_{12}\Delta^*G_{12}\nonumber\\
G_{21}=g_{21}-g_{21}\Delta F^+_{11}-g_{22}\Delta F^+_{21}\nonumber\\
F^+_{21}=g^*_{21}\Delta^*G_{11}+g^*_{22}\Delta^*G_{21}
\label{GrkVx}
\end{eqnarray}
where superconducting energy gap is determined, as usual, from:
\begin{equation}
\Delta^*=VT\sum_{np}F_{11}^+(\varepsilon_n p)=
\lambda T\sum_n\int^{\infty}_{-\infty}
d\xi_pF^+_{11}(\varepsilon_n\xi_p)\equiv\lambda T\sum_n\overline{F^+_{11}
(\varepsilon_n)}
\label{Gap}
\end{equation}
where we have introduced the dimensionless pairing constant 
$\lambda =N_0(0)V$ ($N_0(0)$ -- is the free electron density of states at the
Fermi level.).

The solution of Eqs. (\ref{GrkVx}) gives:
\begin{eqnarray}
G_{11}=-\frac{1}{Det}[(i\varepsilon+\xi_1)(\varepsilon^2+\xi^2_2+D^2+\Delta^2)
-D^2(\xi_1+\xi_2)]=\nonumber\\
=-\frac{1}{Det}\{(i\varepsilon+\xi)[\varepsilon^2
+(\xi+\eta)^2+D^2+\Delta^2]+D^2\eta\}
\label{GF}
\nonumber\\
F^+_{11}=-\frac{1}{Det}\Delta^*(\varepsilon^2+\xi^2_2+D^2+\Delta^2)=
\nonumber\\
=-\frac{1}{Det}\Delta^*[\varepsilon^2+(\xi+\eta)^2+D^2+\Delta^2]
\label{FG}
\end{eqnarray}
where
\begin{eqnarray}
Det=(\varepsilon^2+\xi^2_1+D^2+\Delta^2)(\varepsilon^2+\xi^2_2+D^2+\Delta^2)
-(\xi_1+\xi_2)^2D^2=\nonumber\\
=(\varepsilon^2+\xi^2+D^2+\Delta^2)
(\varepsilon^2+(\xi+\eta)^2+D^2+\Delta^2)-\eta^2D^2
\label{Det}
\end{eqnarray}
where $D$ denotes the real amplitude of fluctuation field (\ref{Vx}).
In accordance with Eq. (\ref{Gap}) Gor'kov's Green's function $F^+_{11}$ 
determines the superconducting energy gap. Taking into account the random 
nature of ``dielectric'' fluctuations Eq. (\ref{Gap}) should be averaged
over the fluctuations of both ``phase'' $\eta=v_Fk$ and amplitude $D$, 
using distributions (\ref{Lork}) and (for high temperature fluctuations) 
(\ref{Rayl}).

Rather long, though direct, calculation of the integral in  (\ref{Gap}) via
residues gives:
\begin{eqnarray}
\overline{F^+_{11}(\varepsilon)}=\frac{\pi\Delta^*}{\sqrt{2}}\frac{1}
{\sqrt{\sqrt{(\tilde\varepsilon^2+D^2+\frac{\eta^2}{4})^2-\eta^2D^2}
+\tilde\varepsilon^2+D^2-\frac{\eta^2}{4}}}
\left\{1+\frac{\tilde\varepsilon^2+D^2+\frac{\eta^2}{4}}
{\sqrt{(\tilde\varepsilon^2+D^2+\frac{\eta^2}{4})^2-\eta^2D^2}}\right\}
\equiv\nonumber\\
\equiv\pi\Delta^*{\cal F}(\varepsilon,\Delta,\eta,D)
\label{ibarF}
\end{eqnarray}
where we have introduced:
\begin{equation}
\tilde\varepsilon=\sqrt{\varepsilon^2+\Delta^2}
\label{tileps}
\end{equation}
Then from Eq. (\ref{Gap}) we can immediately obtain the equation for 
superconducting energy gap in two dimensional ``hot patches'' model
\cite{PS,KS,KS1}:
\begin{equation}
1=2\pi\lambda T\sum_{n=0}^{\left[\frac{\omega_c}{2\pi T}\right]}
\left\{\tilde\alpha{\cal F}(\varepsilon,\Delta,\eta,D)+\frac{1-\tilde\alpha}
{\tilde\varepsilon}\right\}
\label{GapDeta}
\end{equation}
where we have introduced the relative fraction of ``hot'' parts on the Fermi
surface $\tilde\alpha=\frac{4}{\pi}\alpha$. The second term in  (\ref{GapDeta})
gives the standard BCS-like contribution from ``cold'' parts, occupying
$(1-\tilde\alpha)$ part of the Fermi surface. Summation over  $n$ in
(\ref{GapDeta}) is performed up to some maximal value determined by the
integer part of the ratio $\frac{\omega_c}{2\pi T}$.

Numerical solution of (\ref{GapDeta}) allows to find the value of the gap
$\Delta(\eta,D)$ for fixed $\eta$ and $D$ (i.e. for the given value of the
random field of fluctuations (\ref{Vx})) for any given temperature.
After that we can perform averaging over (\ref{Lork}) and (\ref{Rayl}) and find
in this way temperature dependence of the average energy gap. In particular,
for the ``low temperature'' regime of ``dielectric'' fluctuations it is
sufficient to average only over the ``phase'' $\eta$, so that the 
superconducting gap is given by:
\begin{equation}
<\Delta>=\frac{1}{\pi}\int_{-\infty}^{\infty}d\eta\frac{v_F\kappa}
{\eta^2+v_F^2\kappa^2}\Delta(\eta,D)
\label{Gapeta}
\end{equation}
In ``high temperature'' regime we have to add also the averaging over the
amplitude $D$ with distribution function (\ref{Rayl}):
\begin{equation}
<\Delta>=\frac{2}{W^2}\int_{0}^{\infty}dDD\exp\left(-\frac{D^2}{W^2}\right)
\frac{1}{\pi}\int_{-\infty}^{\infty}d\eta\frac{v_F\kappa}
{\eta^2+v_F^2\kappa^2}\Delta(\eta,D)
\label{GapDet}
\end{equation}
As a result we shall find the temperature dependences of the average
superconducting gap $<\Delta>$ without any statistical assumptions like the
self averaging nature of the order parameter. Analogously we can calculate
the temperature dependence of dispersion $<\Delta^2>-<\Delta>^2$, allowing to
estimate the randomness of $\Delta$, 
i.e. the presence or absence of self
averaging. Results of these calculations will be presented in the next section.

As we have already noted in the Introduction, most papers on superconductivity
in disordered systems actually assume the self averaging nature of the
superconducting gap  $\Delta$. In this case $\Delta$
 is treated, in fact, as
non random and independent of parameters of the random field in which the
electrons forming the Cooper pairs actually propagate. In our case these
parameters are the amplitude  $D$ and ``phase'' $\eta$ of (\ref{Vx}), 
accordingly the self averaging over these characteristics may be studied
separately.

Let $\Delta$ be self averaging over the fluctuations of $\eta$. 
In this case we can treat $\Delta$ in Eq. (\ref{FG}) is independent of $\eta$. 
Then the anomalous Gor'kov's function averaged over the fluctuations of
$\eta$ can be written as:
\begin{equation}
<F^+_{11}>=\frac{\Delta^*}{\pi}\int_{-\infty}^{\infty}d\eta
\frac{v_F\kappa}{\eta^2+v_F^2\kappa^2}\frac{\varepsilon^2+(\xi+\eta)^2+D^2
+\Delta^2}{(\varepsilon^2+\xi^2+D^2+\Delta^2)[\varepsilon^2+(\xi+\eta)^2+
D^2+\Delta^2]-\eta^2D^2}
\label{Favph}
\end{equation}
This integral can be directly calculated, so that after long calculations we 
get:
\begin{equation}
<F^+_{11}>=\Delta^*\frac{\tilde\varepsilon^2\left(1+\frac{v_F\kappa}
{\tilde\varepsilon}\right)^2+D^2\left(1+\frac{v_F\kappa}{\tilde\varepsilon}
\right)+\xi^2}
{\left[\left(1+\frac{v_F\kappa}{\tilde\varepsilon}\right)\tilde\varepsilon^2
+\xi^2+D^2\right]^2+v_F^2\kappa^2\xi^2}
\label{Favp}
\end{equation}
where $\tilde\varepsilon$ was introduced in (\ref{tileps}). Accordingly we can
calculate the integral of (\ref{Favp}), entering the gap equation:
\begin{equation}
\overline{<F^+_{11}>}\equiv\int_{-\infty}^{\infty}d\xi <F^+_{11}>=
\frac{\pi\Delta^*\left(1+\frac{v_F\kappa}{2\tilde\varepsilon}\right)}
{\sqrt{D^2+\tilde\varepsilon^2\left(1+\frac{v_F\kappa}{2\tilde\varepsilon}
\right)^2}}
\label{F11av}
\end{equation}
So, despite rather complicated form of the anomalous Green's function
(\ref{Favp}), in the gap equation the account of the interaction with
fluctuations on ``hot'' (flat) parts of the Fermi surface reduces to more or
less ``standard'' renormalization:
\begin{eqnarray}
\varepsilon\to\varepsilon\left(1+\frac{v_F\kappa}{2\tilde\varepsilon}\right)
=\varepsilon\left(1+\frac{v_F\kappa}{2\sqrt{\varepsilon^2+\Delta^2}}\right)
\nonumber\\
\Delta\to\Delta\left(1+\frac{v_F\kappa}{2\tilde\varepsilon}\right)
=\Delta\left(1+\frac{v_F\kappa}{2\sqrt{\varepsilon^2+\Delta^2}}\right)
\label{etaepsD}
\end{eqnarray}
analogous to that appearing in the problem of impurity scattering in
superconductors \cite{AGD}. Similar renormalization was already noted for the
variant of the present problem in Ref. \cite{KS1}. The analogy with impurity
problem here is rather natural as our parameter $v_F\kappa=v_F\xi_{corr}^{-1}$ 
actually determines the characteristic inverse time of flight of an electron
through the region of short range order of the size of
 $\sim\xi_{corr}$. 
Of course, the additional pseudogap influence is also connected with the 
appearance in Eqs. (\ref{Favp}), (\ref{F11av}) of the square of dielectric gap
$D^2$.

Finally, the gap equation determining superconductivity in the ``hot patches''
model with the assumption of self averaging over ``phase'' fluctuations takes
the following form:
\begin{equation}
1=2\pi\lambda T\sum_{n=0}^{\left[\frac{\omega_c}{2\pi T}\right]}
\left\{\tilde\alpha\frac{1+\frac{v_F\kappa}{2\tilde\varepsilon}}
{\sqrt{D^2+\tilde\varepsilon^2\left(1+\frac{v_F\kappa}{2\tilde\varepsilon}
\right)^2}}+\frac{1-\tilde\alpha}
{\tilde\varepsilon}\right\}
\label{Gapfas}
\end{equation}
The solution of this equations, naturally, is simpler than that of Eq.
(\ref{GapDeta})
 with later averaging over (\ref{Gapeta}). In the absence of
fluctuations of the amplitude of ``dielectric'' field $D$, which is valid for
the ``low temperature'' region of fluctuations of short range order, it is
Eq. (\ref{Gapfas}) that determines ``mean field'' (in terms of Ref. \cite{KS})
behavior of $\Delta(T)$ with respect to fluctuations of the random field
(\ref{Vx}).

In ``high temperature'' region of fluctuations of short range order, assuming
for $D$ the distribution function (\ref{Rayl}) and self averaging over the
fluctuations of $D$, we obtain the following equation for the average
superconducting gap:
\begin{equation}
1=2\pi\lambda T\sum_{n=0}^{\left[\frac{\omega_c}{2\pi T}\right]}
\left\{\frac{2\tilde\alpha}{W^2}\int_{0}^{\infty}dDD\exp\left(-\frac{D^2}{W^2}\right)
\frac{1+\frac{v_F\kappa}{2\tilde\varepsilon}}
{\sqrt{D^2+\tilde\varepsilon^2\left(1+\frac{v_F\kappa}{2\tilde\varepsilon}
\right)^2}}+\frac{1-\tilde\alpha}
{\tilde\varepsilon}\right\}
\label{GapfasD}
\end{equation}
describing the situation analogous to that studied in our previous work
\cite{KS1}, where we have analyzed the influence of Gaussian fluctuations of
``dielectric'' short range order using the methods of Refs. \cite{S79,S91}. 
Here the fluctuations of (\ref{Vx})
 are treated exactly, but $D$ is assumed
to be self averaging. We shall see below that the results following from the
solution of Eq. (\ref{GapfasD}) are very close to that obtained in Ref.
\cite{KS1}. For $\kappa\to 0 (\xi_{corr}\to\infty)$ Eq. (\ref{GapfasD})
reduces to the similar ``mean field'' equation of Ref. \cite{KS}.
The temperature of superconducting transition, determined by Eq. (\ref{Gapfas}) 
or Eq. (\ref{GapfasD}), can apparently be identified as the temperature of
the appearance of infinitesimally small superconducting gap, homogeneous over
the whole sample
 \cite{KS}.

In the next section we shall present the results of numerical solution of
Eqs. (\ref{Gapfas}), (\ref{GapfasD}) in comparison with the results of an
exact analysis, based upon the approach using Eqs. (\ref{GapDeta}), 
(\ref{Gapeta}), (\ref{GapDet}).

\section{Main Results and Discussion.}

Let us discuss the results of numerical analysis of gap equations derived in the
previous section\footnote{During our calculations we have assumed for the 
relative part of flat patches on the Fermi the value of $\tilde\alpha = 2/3$}.

\begin{figure}
\epsfxsize=14cm
\epsfysize=14cm
\epsfbox{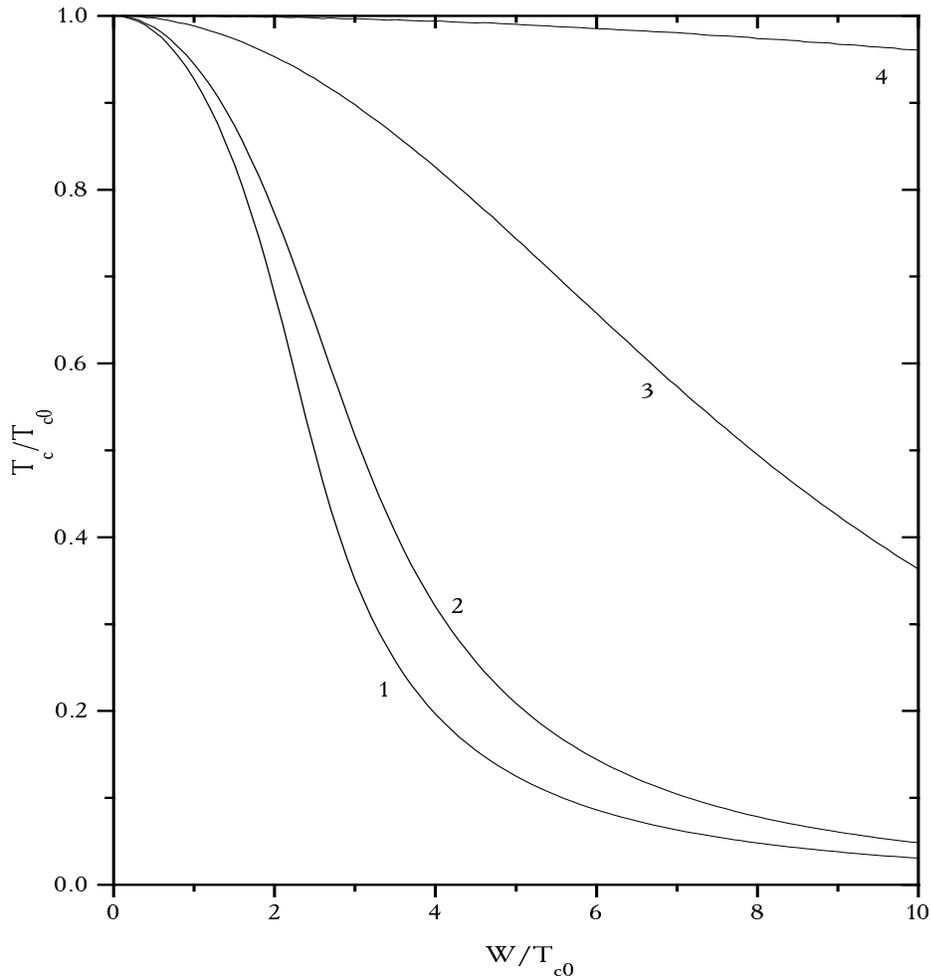}
\caption{Dependence of the critical temperature of superconducting 
transition for the ``low temperature'' region of ``dielectric'' fluctuations
on the width of the pseudogap $W$ for different values of correlation length
of short range order
$\frac{v_F\kappa}{T_{c0}}=$: 1. 0.1; 2. 1; 3. 10; 4. 100.}
\label{figtcw}
\end{figure}

\begin{figure}
\epsfxsize=14cm
\epsfysize=12cm
\epsfbox{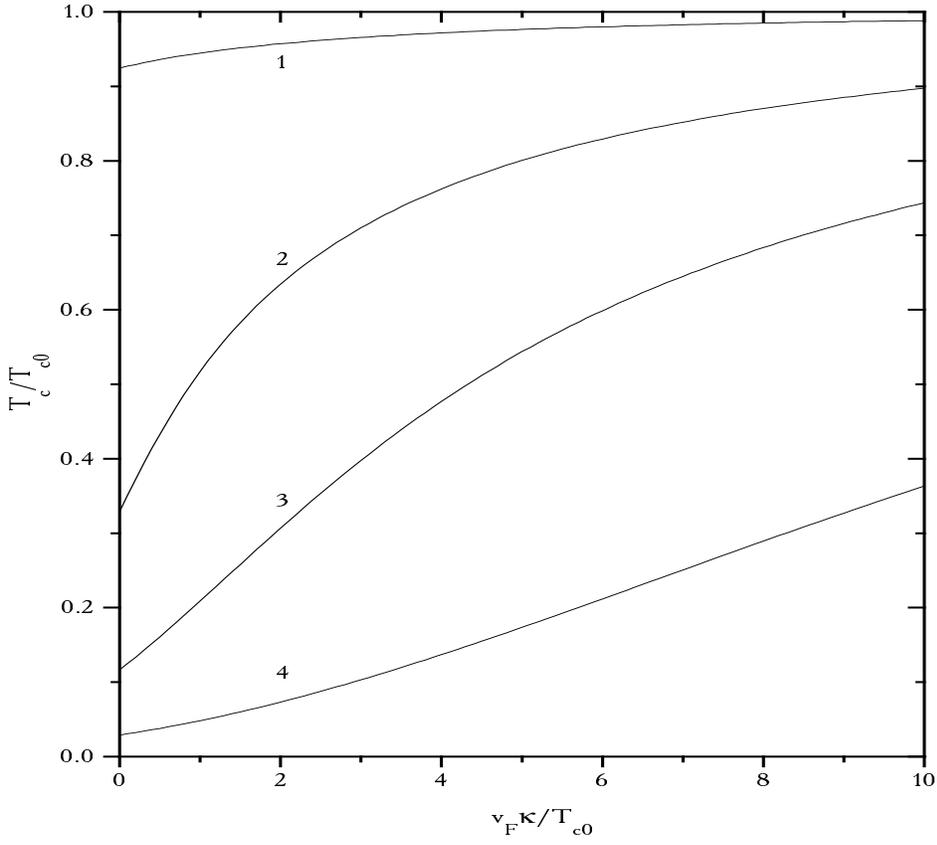}
\caption{Dependence of the critical temperature of superconducting transition
for the ``low temperature'' region of ``dielectric'' fluctuations on the
correlation length of short range order for different values of the
pseudogap width $\frac{W}{T_{c0}}=$: 1. 1; 2. 3; 3. 5; 4. 10.}
\label{figtck}
\end{figure}

In Figs. \ref{figtcw} and \ref{figtck} we show the dependences of the critical
temperature $T_c$ of superconducting transition for the ``low temperature''
regime of ``dielectric'' fluctuations (at this temperature the ``mean field''
gap, determined by Eq. (\ref{Gapfas}) becomes zero)
 on the width of the 
pseudogap  $W$ (which in this case coincide with the amplitude of dielectric
gap $D$) and correlation length of short range order. These results are in
qualitative agreement with similar dependences for the ``high temperature''
regime of ``dielectric'' fluctuations (when $T_c$ is determined by 
Eq. (\ref{GapfasD})),
 as well as with dependences, obtained by us earlier for
the different model of ``dielectric'' fluctuations with finite correlation
length in Ref. \cite{KS1}. With the growth of the pseudogap width $W$ the
``mean field'' value of $T_c$ is suppressed. Diminishing correlation length
leads to the ``filling'' of the pseudogap \cite{MS,S79,BK} and $T_c$ suppression 
becomes less effective.

\begin{figure}
\epsfxsize=14cm
\epsfysize=12cm
\epsfbox{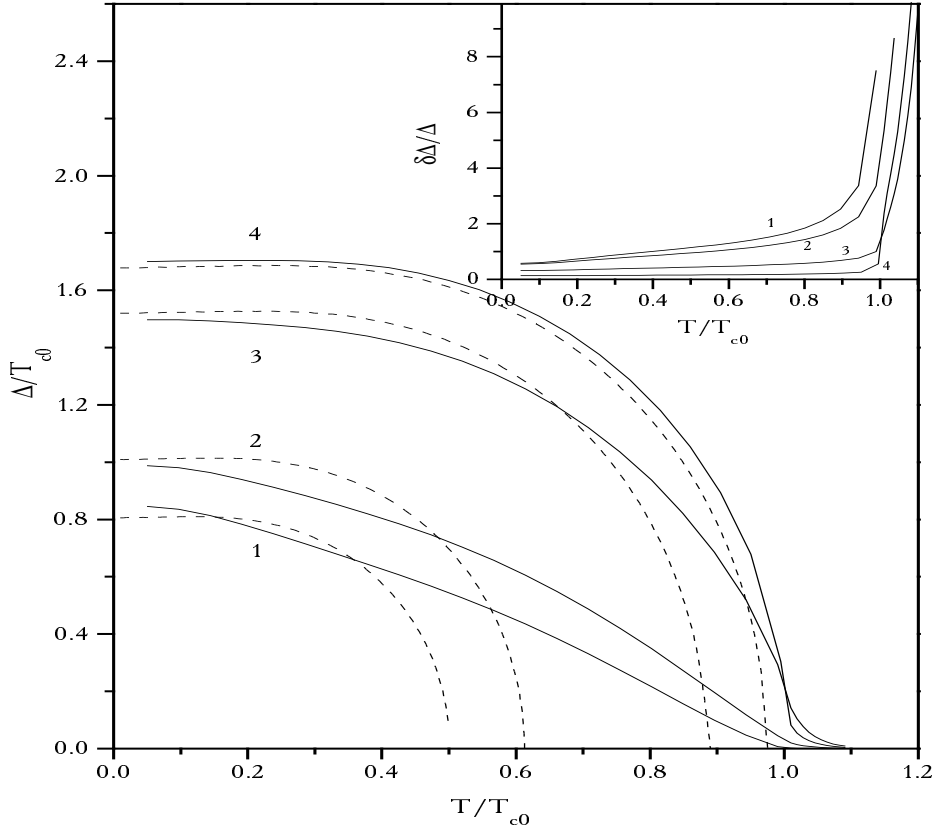}
\caption{Temperature dependence of superconducting energy gap for the
``high temperature'' region of ``dielectric'' fluctuations.
Full curves -- superconducting gap $<\Delta>$ averaged over the amplitude $D$ 
and over the ``phase'' $\eta$, as  determined by Eq. (22).
Dashed curves -- ``mean field'' superconducting gap determined by Eq. (28).
At the insert -- temperature dependence of the relative mean square
fluctuation of superconducting gap. Curves are given for
$\frac{W}{T_{c0}}=3$ and 
$\frac{v_F\kappa}{T_{c0}}=$: 1. 0.1; 2. 1; 3. 10; 4. 100.}
\label{figgt}
\end{figure}

In Fig. \ref{figgt} the curves show the temperature dependences of
superconducting gap $<\Delta>$, averaged over both amplitude  $D$ and ``phase''
$\eta$ (``high temperature'' region of fluctuations of short range order, 
where $<\Delta>$ 
is given by (\ref{GapDet})), for different values of 
$v_F\kappa$. 
Dashed curves give the appropriate ``mean field'' temperature
dependences of superconducting gap, obtained assuming the self averaging of 
superconducting order parameter over both the fluctuations of amplitude and
phase, as  determined by Eq. (\ref{GapfasD}).

\begin{figure}
\epsfxsize=14cm
\epsfysize=12cm
\epsfbox{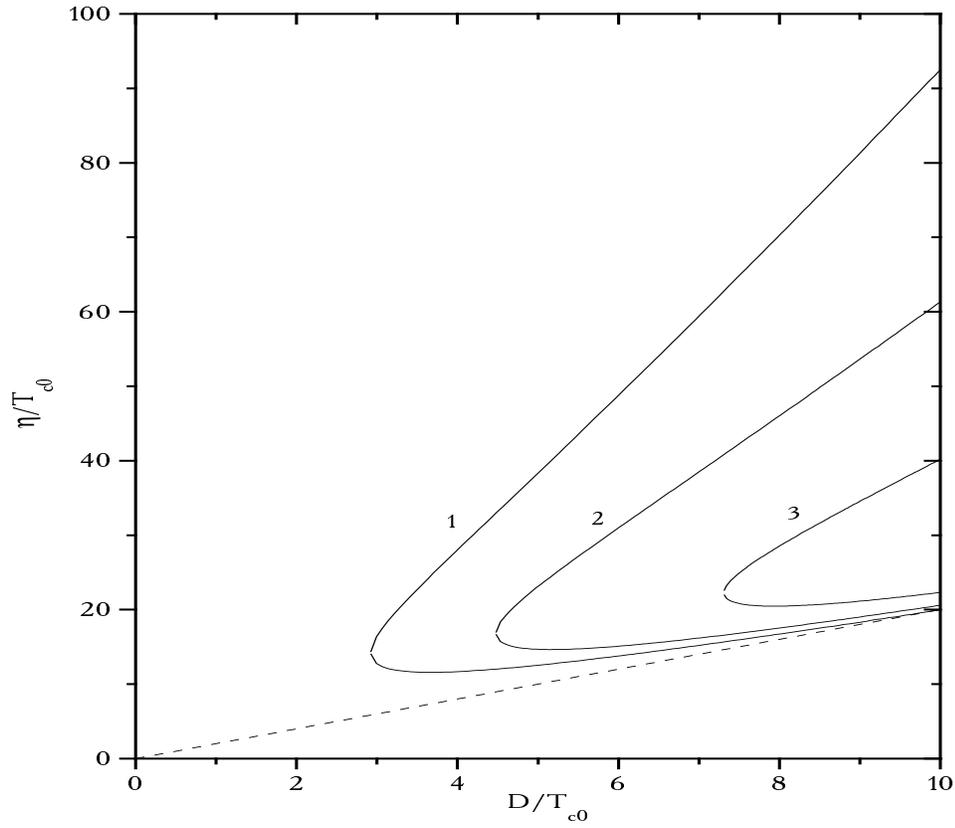}
\caption{The regions of the phase diagram with non zero superconducting
gap for different temperatures above $T_{c0}$, shown for the values of
$T/T_{c0}$: 1. 1.05; 2. 1.1; 3. 1.2. 
Dashed line corresponds to $D=\eta /2$.}
\label{fignd}
\end{figure}

Superconducting gap, averaged over the fluctuations of short range order,
is non zero in the temperature region above $T_c$, corresponding to the zero
of the ``mean field'' gap (i.e. the gap homogeneous over the sample).
More so, it is seen that the average gap is non zero also in a narrow region
of temperatures larger than the transition temperature in the absence of
short range order fluctuations (pseudogap) $T_{c0}$. 
This effect is due to the
existence of fluctuations of the ``phase'' $\eta$, when the Fermi level is
in the energy interval, corresponding to the peaks of the density of states
due to formation of dielectric gap. To understand this, note that for the
given realization of ``phase'' $\eta$ and of amplitude $D$, the density of
states has the form: 
\begin{equation}
\frac{N(E)}{N_0(0)}=-\frac{1}{\pi N_0(0)}Im\sum_{\bf p}g_{11}^{R}(Epp)=
\left\{
\begin{array}{l}
\frac{|E+\frac{\eta}{2}|}{\sqrt{(E+\frac{\eta}{2})^2-D^2}}\qquad\mbox{for}
\quad |E+\frac{\eta}{2}|>D\\
0\qquad\mbox{for other values}
\end{array}
\right.
\label{DosD}
\end{equation}
where $g_{11}^{R}(Epp)$ -- is the retarded Green's function, obtained from
(\ref{g11}) via standard analytical continuation $i\varepsilon_n\to E+i0$,
and $N_0(0)$ -- is the density of states at the Fermi level in the absence
of short range order fluctuations. Thus for $\frac{\eta}{2}\approx D$ the Fermi
level position approximately coincides with the peak of the density of states,
leading to larger values of superconducting gap $\Delta(\eta,D)$. The growth 
of the dielectric gap $D$
 leads also to the larger width of peaks in the 
density of states (\ref{DosD}), 
so that for $\frac{\eta}{2}\approx D$ 
superconducting gap $\Delta(\eta,D)$ grows with $D$. Then, for any temperature
above $T_{c0}$ and for large enough amplitudes of dielectric gap 
$D>D^{\ast}(T)$, on the phase diagram in the variables $\eta$ and $D$ there is
always a narrow region close to the line 
$\frac{\eta}{2}=D$, where
superconducting gap $\Delta(\eta,D)$ is different from zero 
(cf. Fig.\ref{fignd}). This leads to the appearance in the temperature
dependence of the averaged (over random field fluctuations) superconducting gap
$<\Delta >$ of an infinite, though exponentially small ``tail'' in the 
temperature region above $T_{c0}$.

At the insert in Fig. \ref{figgt} we show the temperature dependence of the
relative mean square fluctuation of the superconducting gap
$\delta\Delta/\Delta=\sqrt{<\Delta^2>-<\Delta>^2}/<\Delta>$
 for the
``high temperature'' regime of ``dielectric'' fluctuations. For large
correlation lengths of short range order ($\xi_{0}/\xi_{corr}\ll 1$) 
these
fluctuations of superconducting order parameter are very strong for all
temperatures, signifying the obvious absence of self averaging. Surprisingly,
these fluctuations of superconducting energy gap are strong enough also for
small enough correlation lengths, at least in the region of $T>T_c$. 
In particular, the ``tail'' in the temperature dependence of
$<\Delta >$ for $T>T_c$ is observed even for  $v_F\kappa /T_{c0}=100$, when 
$\xi_{0}/\xi_{corr}\approx 30\gg 1$.

\begin{figure}
\epsfxsize=14cm
\epsfysize=12cm
\epsfbox{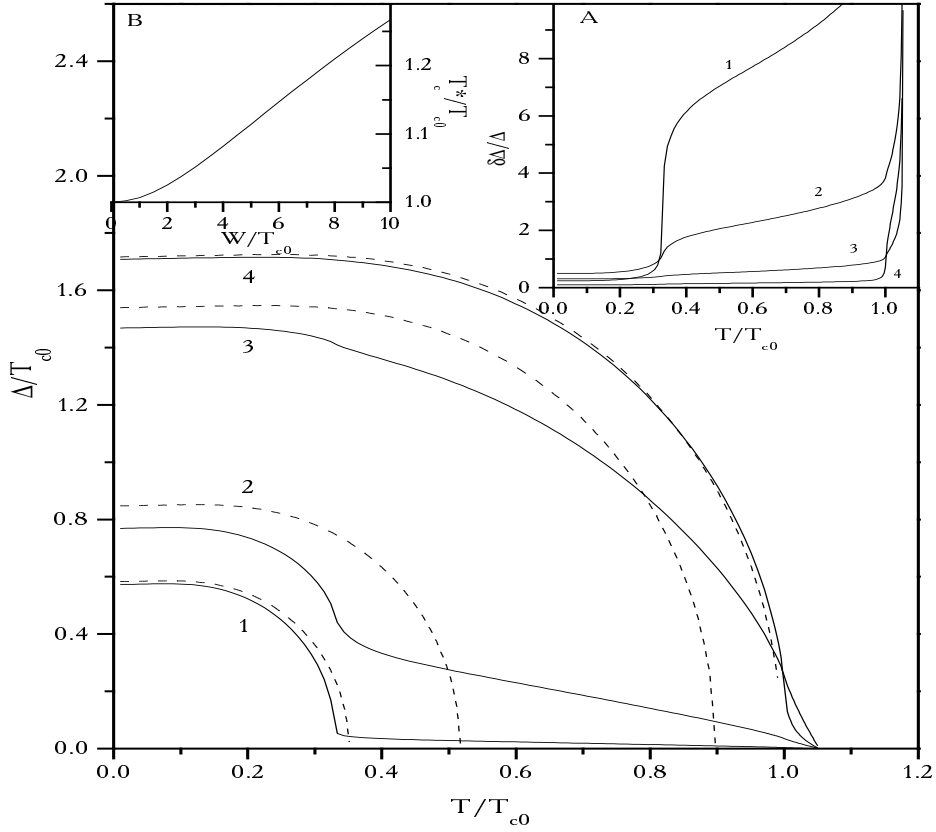}
\caption{Temperature dependence of superconducting gap for the ``low
temperature'' region of ``dielectric'' fluctuations.
 Full curves --
gaps $\Delta$ averaged over ``phase'' $\eta$ for fixed values of the amplitude 
$D=W$
, as determined by Eq. (27). Dashed curves -- ``mean field'' gap,
determined by Eq. (27). At the insert A -- temperature dependence of the
relative fluctuation of superconducting gap. Curves are given for
$\frac{W}{T_{c0}}=3$ and $\frac{v_F\kappa}{T_{c0}}=$: 
1. 0.1; 2. 1; 3. 10; 4. 100.
At the insert B -- dependence of the critical temperature $T_c^\ast$ 
on the width of the pseudogap.}
\label{figgtp}
\end{figure}

The full curves in Fig. \ref{figgtp} show the temperature dependence of
superconducting gap $<\Delta >$, averaged over the ``phase''  $\eta$ 
(cf. (\ref{Gapeta})), for the ``low temperature regime'' of ``dielectric''
fluctuations, when the amplitude fluctuations of dielectric gap are frozen out
and $D=W$. Dashed curves show the appropriate temperature dependences of the
``mean field'' superconducting gap, obtained with the assumption of self
averaging superconducting order parameter over fluctuations of the ``phase''
$\eta$, and defined by Eq. (\ref{Gapfas}).
 For large enough correlation lengths
of the short range order the average gap for $T<T_c$ is very close to its
``mean field'' values and its ``tail'' in the region of $T>T_c$ is relatively
small. This behavior for the ``low temperature'' region of ``dielectric''
fluctuations is due to the fact, that for 
 $\xi_{corr}\to \infty$ there is no
randomness in this model at all ($\eta =0$, $D=W$). Accordingly, the mean
square fluctuation of the gap, shown at the insert in Fig. \ref{figgtp}, 
is rather small for large correlation lengths and $T<T_c$, but grows sharply
for $T>T_c$. As correlation length becomes smaller, fluctuations of
superconducting gap $\delta\Delta$ for $T<T_c$ at first grow
, mainly due to
the growth of randomness (parameter $v_F\kappa$ 
determines the width of the
distribution of fluctuations of $\eta$), but afterwards diminish in the region 
of $\xi_{0}/\xi_{corr}\gg 1$. In the ``tail'' region of the averaged 
superconducting gap ($T>T_c$) fluctuations of superconducting gap are quite
large. Though these fluctuations diminish for smaller correlation lengths of
the short range order $\xi_{corr}$, 
they still remain significant even for
small enough correlation lengths, i.e. for $\xi_{0}/\xi_{corr}\gg 1$.

As well as in the ``high temperature'' region of ``dielectric'' fluctuations
the ``tail'' in the temperature dependence of the average gap is observed here
also in the region of $T>T_{c0}$. It is explained by the same reasons as 
discussed above. However, for the ``low temperature'' region the amplitude
of the dielectric gap is non random and fixed at $D=W$.
 Thus, for
$T_{c0}<T<T_c^\ast$, where $T_c^\ast$ is defined by $D^\ast(T_c^\ast )=W$, 
there exists a narrow region of ``phases'' close to $\eta =2W$, where the
superconducting gap $\Delta (\eta ,W)$ is different from zero, while for
$T>T_c^\ast$
 such region is absent (cf. Fig.\ref{fignd}). The value of
$T_c^\ast$ determines the temperature up to which exists the ``tail'' of
the average gap, i.e. the critical temperature for the average gap 
 $<\Delta >$. 
From the definition of $T_c^\ast$ it is obvious that this temperature is 
independent of correlation lengths and depends only on $W$. As the width of the
peaks in the density of states 
(\ref{DosD}), as well as $\Delta (\eta ,D)$,
grows with the growth of $D$ if condition $\frac{\eta}{2}\approx D$ is 
fulfilled, 
the value of $T_c^\ast$ grows with $W$. The dependence of
$T_c^\ast$ on $W$ is shown at the insert in Fig. \ref{figgtp}.

\section{Conclusion}

In this work we have studied superconductivity within very simple model of the
pseudogap in two dimensional electronic system which allows an exact solution.
The central result is the explicit demonstration of the absence[21~e of complete
self averaging of superconducting order parameter (energy gap) over the
random field of ``dielectric'' fluctuations, leading to the formation of the
pseudogap. This is rather surprising from the point of view of the standard
theory of superconductivity in disordered systems \cite{Gor,Genn,SL}. 
The absence of self averaging is reflected in the appearance of 
fluctuations of the gap, especially strong for the temperatures larger than
the ``mean field'' temperature of superconducting transition $T_c$, following
from the standard equations, assuming the the self averaging nature of the
order parameter. We identify this temperature with the critical temperature
for the appearance of homogeneous superconducting state in the whole sample,
while for the real disordered system the superconducting state is 
inhomogeneous and for $T>T_c$ 
superconductivity exists in separate regions
(``drops''), appearing due to random fluctuations of the local density of
electronic states. The difference with our previous work \cite{KS}, 
where this picture was discussed in the limit of very large correlation 
lengths of short range order $\xi_{corr}\to\infty$, the use of the model of
Ref. \cite{BK} allowed to find the complete solution for arbitrary values of
$\xi_{corr}$. This solution demonstrates the absence of complete self
averaging of superconducting gap even for
 $\xi_{corr}<\xi_{0}$, which
contradicts the naive expectations of the standard approach \cite{MS}. 
As we noted above, we are unaware of works, where the problem of
self averaging of $\Delta$ was analyzed within any exactly solvable model of
disorder. Here we presented precisely this type of analysis.
It is unclear at present whether the results obtained will be conserved in
more realistic models of disorder.

It will be very interesting to analyze also the behavior of the spectral
density and tunneling density of states, analogous to that done in our
previous work \cite{KS} in the limit of $\xi_{corr}\to\infty$. It will be
especially important to consider the self averaging properties of the density
of states, which is the standard assumption in the theory of disordered 
systems.
 
As to comparison with experiments on high -- temperature superconductors,
we note Refs. \cite{Cren1,Cren2}, where the scanning tunneling microscopy
(measuring the local density of states) on films of
$Bi_2Sr_2CaCu_2O_{8+\delta}$ clearly demonstrated the existence in this system
of microscopic superconducting regions along with dominant semiconducting
regions with typical pseudogap in the electronic spectrum. These observations
are in qualitative agreement with the main conclusions of the present study.

This work was partly supported by the grants 99-02-16285 of the Russian
Foundation of Basic Research and CRDF No. REC-005, as well as by the program of
fundamental research of the Presidium of the Russian Academy of Sciences
``Quantum macrophysics'', and by the State Contracts of the Ministry of
Industry and Science 108-11(00) (program ``Statistical Physics'') and  
107-1(00) (program of HTSC research).

\newpage

\newpage

\newpage

\newpage

\newpage

\newpage

\newpage
\begin{references}
\bibitem{Tim}T.Timusk, B.Statt. Rep.Progr.Phys. {\bf 62}, 61 (1999)
\bibitem{MS}M.V.Sadovskii. Uspekhi Fiz. Nauk {\bf 171}, 539 (2001); 
Physics Uspekhi {\bf 44}, 515 (2001); ArXiv: cond-mat/0102111
\bibitem{Lok}V.M.Loktev, R.M.Quick, S.G.Sharapov. Phys. Reports {\bf 349}, 2 
(2001); ArXiv: cond-mat/0012082 
\bibitem{PS}A.I.Posazhennikova, M.V.Sadovskii. Zh.Eksp.Teor.Fiz. (JETP) 
{\bf 115}, 632 (1999); ArXiv: cond-mat/9806199
\bibitem{KS}E.Z.Kuchinskii, M.V.Sadovskii. Zh.Eksp.Teor.Fiz. (JETP)
{\bf 117}, 613 (2000); ArXiv: cond-mat/9910261; 
Physica {\bf C341-348}, 879 (2000)
\bibitem{KS1}E.Z.Kuchinskii, M.V.Sadovskii. Zh.Eksp.Teor.Fiz. (JETP)
{\bf 119}, 553 (2001); ArXiv: cond-mat/0008377
\bibitem{S74} M.V.Sadovskii. Zh.Eksp.Teor.Fiz. (JETP) {\bf 66},1720(1974); 
Fiz.Tverd.Tela (Sov.Phys. - Solid State) {\bf 16},2504(1974) 
\bibitem{S79}M.V.Sadovskii. Zh.Eksp.Teor.Fiz. (JETP) {\bf 77}, 2070(1979) 
\bibitem{S91}M.V.Sadovskii, A.A.Timofeev. Superconductivity: Physics, Chemistry,
Technology {\bf 4}, 11(1991); J.Moscow Phys.Soc. {\bf 1},391(1991)
\bibitem{Sch} J.Schmalian, D.Pines, B.Stojkovic. Phys.Rev.Lett. 
{\bf 80}, 3839(1998); Phys.Rev. {\bf B60}, 667 (1999); ArXiv: cond-mat/9804129
\bibitem{KS2}E.Z.Kuchinskii, M.V.Sadovskii. Zh.Eksp.Teor.Fiz. (JETP) {\bf 115}, 
1765 (1999); ArXiv: cond-mat/9808321
\bibitem{Gor}L.P.Gor'kov. Zh.Eksp.Teor.Fiz. (JETP) {\bf 37}, 1407 (1959)
\bibitem{Genn}P.De Gennes. Superconductivity of Metals and Alloys. 
W.A.Benjamin, NY, 1966
\bibitem{SL}M.V.Sadovskii. Superconductivity and Localization. World
Scientific, Singapore, 2000;  Phys.Reports {\bf 282}, 225 (1997);
ArXiv: cond-mat/9308018; Superconductivity: Physics, Chemistry, Technology
{\bf 8}, 337 (1995);
\bibitem{BK}L.Bartosch, P.Kopietz. Eur. Phys. J. {\bf B17}, 555 (2000);
ArXiv: cond-mat/0006346
\bibitem{LRA}P.A.Lee, T.M.Rice, P.W.Anderson. Phys.Rev.Lett. {\bf 31}, 462
(1973)
\bibitem{Sad}M.V.Sadovskii. Physica {\bf C341-348}, 811 (2000);
ArXiv: cond-mat/9912318
\bibitem{B1}L.Bartosch, P.Kopietz. Phys.Rev. {\bf B60}, 15488 (1999);
ArXiv: cond-mat/9908065
\bibitem{B2}L.Bartosch. Ann. der Physik (in press); ArXiv: cond-mat/0102160
\bibitem{MM}A.J.Millis, H.Monien. Phys.Rev. {\bf B61}, 12496 (2000);
ArXiv: cond-mat/9907223
\bibitem{BD}S.A.Brazovskii, I.E.Dzyaloshinskii. 
Zh.Eksp.Teor.Fiz. (JETP) {\bf 71}, 2338 (1976)
\bibitem{Str}J.Tranquada. J.Phys.Chem.Sol. {\bf 59}, 2150 (1998);
ArXiv: cond-mat/9802043
\bibitem{AGD}A.A.Abrikosov, L.P.Gor'kov, I.E.Dzyaloshinskii. 
Methods of the Quantum Field Theory in Statistical Physics.
Prentice-Hall, NY, 1963
\bibitem{Cren1}T.Cren, D.Roditchev, W.Sacks, J.Klein, J.-B.Moussy,
C.Deville-Cavellin, M.Lagues. Phys.Rev.Lett. {\bf 84}, 147 (2000)
\bibitem{Cren2}T.Cren, D.Roditchev, W.Sacks, J.Klein. Europhys. Lett. (in press)
\end{references}
\end{document}